\documentclass[pdfusetitle,11pt,a4paper]{article}
\usepackage{amsmath}
\usepackage[dvipdfmx,hiresbb]{graphicx}
\usepackage{xcolor}
\usepackage{mathrsfs,mathtools}
\usepackage{physics,amssymb}
\usepackage{bm}
\usepackage{braket}
\usepackage{listings}

\usepackage{cases}
\usepackage{soul}
\usepackage{cancel}
\usepackage{cases}
\usepackage[utf8]{inputenc}
\usepackage{url}
\usepackage{float}
\usepackage{longtable}
\usepackage[normalem]{ulem}
\usepackage{xspace}
\usepackage{aas_macros}
\usepackage{autobreak}

\definecolor{Green}{HTML}{0B6623}

\usepackage{ulem}
\DeclareRobustCommand{\Erase}{\bgroup\markoverwith{\textcolor{red}{\rule[.5ex]{2pt}{0.4pt}}}\ULon}

\usepackage{subcaption}
\usepackage{empheq}
\usepackage{here}
\usepackage{pdfpages}

\usepackage{multirow}

\usepackage{tikz}
\usetikzlibrary{intersections,calc,arrows.meta,decorations.markings,decorations.pathmorphing}
\tikzset{snake it/.style={decorate, decoration=snake}}
\setstcolor{red}
\usepackage{jcappub}
\hypersetup{colorlinks=true
,urlcolor=DARKBLUE
,anchorcolor=DARKBLUE
,citecolor=DARKBLUE
,filecolor=DARKBLUE
,linkcolor=DARKBLUE
,menucolor=DARKBLUE
,linktocpage=true
,pdfproducer=medialab
}

\usepackage{amsthm}
\theoremstyle{definition}

\newtheorem*{prop*}{Proposition}

\definecolor{MONZA}{HTML}{CF000F}
\definecolor{DARKBLUE}{HTML}{00008b}
 \definecolor{DARKBLUE}{rgb}{0,0,0.7} 
\definecolor{DARKMAGENTA}{HTML}{8b008b}
\definecolor{DARKCYAN}{HTML}{008B8B}
\definecolor{DARKORANGE}{HTML}{FF8C00}

\definecolor{MAGENTA}{HTML}{FF00FF}

\begin{document}

\title{Primordial black hole formation from a type II perturbation in the absence and presence of pressure}

\author[a]{Koichiro Uehara,}
\author[a,b]{Albert Escriv\`{a},}
\author[c]{Tomohiro Harada,}
\author[d]{~~~~~~~~~~~Daiki Saito,}
\author[a,e]{and Chul-Moon Yoo}

\affiliation[a]{Division of Science, Graduate School of Science, Nagoya University, Nagoya 464-8602, Japan}
\affiliation[b]{Institute for Advanced Research, Nagoya University, Furo-cho Chikusa-ku, Nagoya 464-8601, Japan}
\affiliation[c]{Department of Physics, Rikkyo University, Toshima, Tokyo 171-8501, Japan}
\affiliation[d]{Department of Physics, Kyoto University, Kyoto 606-8502, Japan}
\affiliation[e]{Kobayashi Maskawa Institute,
Nagoya University, Nagoya 464-8602, Japan}

\emailAdd{uehara.koichiro.p8@s.mail.nagoya-u.ac.jp}
\emailAdd{escriva.manas.alberto.k0@f.mail.nagoya-u.ac.jp}
\emailAdd{harada@rikkyo.ac.jp}
\emailAdd{saito@tap.scphys.kyoto-u.ac.jp}
\emailAdd{yoo.chulmoon.k6@f.mail.nagoya-u.ac.jp}

\date{\today}
\abstract{
We investigate primordial black holes (PBHs) formed from extremely large amplitudes of primordial curvature fluctuations, classified as type II.
Type II fluctuations differ from type I by the presence of a stationary point on the initial time slice, when we see the areal radius as a function of the radial coordinate.
Starting from these type II perturbations to form black holes, the nonlinear evolution governed by the Einstein equations generally results in two distinct types, A and B, of horizon configurations, respectively characterized by the absence and presence of a bifurcating trapping horizon where past and future trapping horizons meet.
In this paper, we use the Lema\^itre--Tolman--Bondi solution to show that type I/II and type A/B classifications are equivalent for a spherically symmetric dust fluid system, regardless of the fluctuation profile. 
However, this equivalence does not generally hold in the presence of pressure.
}

\arxivnumber{2505.XXXXX}

\maketitle
\flushbottom
\section{Introduction}
The dynamics of gravitational collapse is essential for understanding the formation process of celestial bodies. 
Newtonian gravity has been used to analyze gravitational collapse leading to the formation of nonrelativistic stars. 
Since the advent of general relativity, research into gravitational collapse and its final fate, including the formation of black holes, has been at the forefront of this field.
One of the simplest analytic examples in the general relativistic approach 
is the Oppenheimer--Snyder collapse~\cite{Oppenheimer:1939ue}.
In general, numerical simulations are necessary to investigate more realistic situations, such as stellar collapse or compact binary coalescence.

While black holes are often linked to astrophysical processes, such as stellar evolution, they can also arise from cosmological phenomena, including the collapse of primordial density perturbations, which leads to the formation of primordial black holes (PBHs).
The standard scenario of PBH formation is described as the gravitational collapse of an overdense region in the radiation-dominated flat Friedmann--Lema\^itre--Robertson--Walker (FLRW) background universe~\cite{Zeldovich:1967lct,10.1093/mnras/152.1.75,10.1093/mnras/168.2.399, Carr:1975qj}. 
The collapse occurs if self-gravity overcomes pressure gradient, that is, the amplitude of the density perturbation $\delta$ exceeds a certain threshold value, which generally depends on the specific shape of the density perturbation profile and the equation of state of the background universe. 

In the context of the PBH formation, there was a long-standing issue with the large density fluctuation limit. 
Since the local dense region can be regarded as a part of a closed universe, it would be completely closed and lose the possible boundary 
connected to the background universe in the limit of the large amplitude of the density fluctuation~\cite{10.1093/mnras/168.2.399, PhysRevD.71.104009}.
This situation has sometimes been called a separate universe, 
and it is often excluded from realizing PBH formation in the universe without a concrete understanding. 
A resolution to this issue was proposed by Kopp, Hofmann, and Weller (KHW)~\cite{PhysRevD.83.124025}, who re-derived the upper bound on the density perturbation based on the spatially hemispherical geometry of an initial time slice, rather than by referring to the separate universe scale. 
Their analysis revealed that the separate universe condition corresponds to the limit at which the curvature perturbation diverges.
Accordingly, it is not problematic when one considers the curvature fluctuation.

They also introduced the classification of ``type I'' and ``type II'' initial perturbations depending on the behavior of the areal radius $R$ as a function of the radial coordinate $r$ on the initial time slice. 
If the areal radius on a spacelike hypersurface of constant $t$ with the limit $t\to 0$ as sufficiently close to the big bang is non-monotonic as a function of the radial coordinate $r$, the initial perturbation is classified into ``type II'', while a monotonic one is ``type I''. 
In other words, ``type II'' fluctuation is equivalently defined as exhibiting non-singular extremal point $\{r_p\}$ of areal radius $R(r)$ for a coordinate radius $r$, called having ``necks'' and ``bellies'', as local minimum and maximum points, respectively.\footnote{
Also, this is simply called a ``bottle neck'' or ``wormhole'', as a generalization of the Kruskal--Szekeres throat or the Einstein--Rosen bridge at $r=2m$ of the Schwarzschild solution, cf. Refs.~\cite{osti_4173295,A_Barnes_1970,10.1093/mnras/207.1.23P,10.1143/PTP.72.1137,1985GReGr..17..251S,Hellaby:1985zz,Hellaby1987,PhysRevD.69.043502,Hellaby:2002nx} or textbook~\cite{plebanski2006introduction}.
}
\footnote{
In this paper, we assume regularity at those points and do not consider a singularity.
Although in the dust collapse model the extrema can be associated with a shell-crossing singularity, we do not consider this case.
This is why the structure is defined as ``non-singular'' extrema.
}
Regarding the curvature fluctuation $\zeta$, the areal radius is expressed as $R\propto r \exp \zeta$.
Then, if the amplitude is sufficiently large, ``type II'' fluctuation, i.e., non-monotonicity, is realized. 
Although the probability for type II perturbation is usually expected to be smaller than type I, both contribute to the PBH abundance, and in specific models, the type II contribution can be larger than that of type I~\cite{Gow_2023, Escriva:2023uko, Shimada:2024eec, Inui:2024fgk, Escriva:2025rja} (see also \cite{Deng:2016vzb, Deng:2017uwc} for other scenarios of PBH formation from domain walls and vacuum bubbles, where a throat structure in the areal radius emerges during the time evolution).

KHW also analyzed the dynamics of PBH formation resulting from ``type I'' and ``type II'' perturbations, as well as the difference between them, using the Lema\^itre--Tolman--Bondi (LTB) solution. 
They found that, for a model of ``type II'' initial fluctuation, the non-monotonic behavior of the areal radius is preserved during the whole evolution with synchronous time slices in the LTB solution. 
Then, the non-central region pinches off when the neck region touches the future spacelike singularity. 
They observed that the pinch-off behavior of the neck is seen as an apparent topology change when using coordinate time $t$-constant slices.
However, they concluded that this character is a slice-dependent issue and no topology change occurs in the conformal diagram.

Recently, we numerically investigated the dynamics of type II PBH formation, defined by the PBH formation from type II fluctuation, in the radiation-dominated universe~\cite{Uehara:2024yyp}, for what we examined the configuration of the trapping horizons~\cite{Hayward:1993wb, Hayward:1994bu} and classified the PBH formation with/without bifurcating trapping horizons~\cite{Maeda:2009tk} as type B/A as a distinct difference. 
In this work, we simply define type II as the existence of a stationary point of $R$ for $r$, which includes the non-monotonic case and the marginal one.

Then, we observed that the neck structure can disappear through time evolution in the case of a radiation-dominated universe, as KHW had anticipated as a difference from their dust model. 
Consequently, to obtain bifurcating trapping horizons, the initial amplitude of the curvature fluctuation must be significantly larger than the threshold value between types I and II. 
That is, we found the existence of type II-A PBH formation in the radiation-dominated universe.

Furthermore, Refs.~\cite{Shimada:2024eec, Inui:2024fgk} recently identified specific cases of type II perturbations that do not lead to black hole formation in a radiation-dominated universe, considering typical profiles derived with statistical non-Gaussianities in the primordial perturbation. 
On the other hand, Refs.~\cite{Escriva:2025eqc, Escriva:2025rja} examined different perturbation profiles, formulating criteria to identify non-collapsing type II perturbations and deriving an analytical estimate for predicting the black hole formation threshold.

In KHW, only two types, type I-A and II-B, were observed for a specific initial profile of the LTB solution. 
This suggests that type I/II and type A/B are equivalent for the dust fluid system in general, regardless of the profile of the initial fluctuation. 
We clarify that, for the LTB solution, the existence of the neck structure in the initial data of the long-wavelength limit is the necessary and sufficient condition for the existence of a bifurcating trapping horizon. 
We can analytically prove the equivalence between type I/II and type A/B classifications for the PBH formation in a dust-dominated universe. 

This paper is organized as follows.
Section~\ref{sec:BasicEqs.} describes the setup.
In Section~\ref{subsec:CC3+1andLWLapprox.}, we introduce curvature fluctuation in the long-wavelength limit for initial data of PBH formation and define type I/II classification. 
Then, we translate the situation in terms of the LTB spacetime in Section~\ref{subsec:LTBinLWL}.
Section~\ref{subsec:EvolutionInLTB} considers the PBH formation and introduces type A/B spacetimes regarding the trapping horizons.
The main results are given in Section~\ref{sec:NoGap}, in which the equivalence between type I/II and type A/B in the LTB solution is shown.
Section~\ref{subsec:SuffCond} describes the sufficient condition; if a perturbation is of type II, it must result in a type B spacetime.
Section~\ref{subsec:curvatureFluc} shows the necessary condition; if a spacetime is of type B, it must result from a type II perturbation.
Section~\ref{sec:AnalyDemoWithExampleProfile} demonstrates the analysis with an example profile. 
Section~\ref{sec:conclusion} is devoted to a summary.

Throughout the paper, we use the geometrical unit in which the speed of light and the gravitational Newton constant are unity, that is, $c = G =1$.

\section{Basic equations}\label{sec:BasicEqs.}
\subsection{Cosmological conformal 3+1 decomposition and long-wavelength approximation}\label{subsec:CC3+1andLWLapprox.}

In this subsection, we focus on the adiabatic growing modes, supposing that a perfect fluid gives the matter content. 
For a spherically symmetric spacetime, the metric can be expressed using a cosmological conformal 3+1 decomposition as
\begin{align}
    ds^2 &= -\alpha^2 dt^2 + a^2(t) \psi^4(t,r)\qty((dr + \beta^r dt)^2 + r^2 d\Omega^2), \label{eq:CC3+1decomp}
\end{align}
where $\alpha$, $\beta^r$, $\psi$, $a$, and $d\Omega^2$ represent the lapse function, the radial component of the shift vector, the spatial conformal factor, the scale factor from the asymptotic background of a flat FLRW solution, and the line element of a 2-sphere, respectively. 

Introducing the typical scale of the spatial gradient $k$, we can perform the gradient expansion with the non-dimensional expansion parameter $\epsilon \coloneqq k/aH\ll1$, with $H(t) \equiv d \ln a(t) /dt$ being the Hubble factor of the FLRW solution 
(see, e.g., Refs.~\cite{PhysRevD.60.084002,Lyth_2005,Harada:2015ewt}). 
Then, for the adiabatic growing mode, the metric can be expressed as 
\begin{align}
    ds^2 
    &\approx -dt^2 + a^2(t)e^{2\zeta(r)}\qty(dr^2 + r^2d\Omega^2), \label{eq:zetaInMetric}
\end{align}
where the weak equality denotes the equality in the limit $\epsilon\rightarrow 0$. 
This expression shows that the curvature perturbation $\zeta$ encapsulates primordial inhomogeneities, corresponding to the adiabatic growing mode in the long-wavelength approximation~\cite{Tolman:1934za, PhysRevD.60.084002, Lyth_2005, Polnarev:2006aa, PhysRevD.91.084057}.
Once a spatial profile of $\zeta(r)$ and the gauge conditions for the time slicing and threading of spatial coordinates are fixed, all geometrical and matter quantities are analytically determined order by order with the gradient expansion parameter $\epsilon$ up to quadratures. 

To introduce the classification of the primordial fluctuations, first, let us define the areal radius as follows:
\begin{align}
    R(t,r) \coloneqq a(t)\psi^2(t,r)r
    \approx a(t)e^{\zeta(r)}r. 
\end{align}
Following Ref.~\cite{PhysRevD.83.124025}, we classify the primordial fluctuation into type II/I if the areal radius is with/without stationary points of $R$ as a function of the isotropic coordinate $r$ for the initial time slice in the long-wavelength limit. 
Since $\partial_r R\propto \partial_r(r\exp \zeta(r)) = (1+r\partial_r\zeta(r))\exp \zeta(r) $, for a large amplitude of $\zeta$, the value of $r\partial_r\zeta(r)$ can take a negatively larger value than $-1$, and the type II fluctuation can be realized with $\partial_rR=0$ at certain radii $r=r_p$. 
In this paper, we do not use the original definition of ``type II'' fluctuation via non-monotonicity of $R(r)$ for $r$ for $t\to0$ limit.
We redefine \emph{type II} just by the existence of a stationary point $r=r_p$, satisfying 
\begin{align}
    \partial_rR(r_p) = 0,
\end{align}
for the long wavelength solution at the $t\to 0$ limit, while positive definite $\partial_rR > 0$ implies \emph{type I}.
Thus, type II fluctuations, as defined here, contain the ones of non-monotonic $R(r)$, which are categorized as ``type II'' with the original definition in KHW.
Specific examples will be explicitly shown in the following sections.

\subsection{The LTB solution in the long-wavelength limit}\label{subsec:LTBinLWL}

The LTB solution~\cite{Lemaitre:1933gd, Tolman:1934za} (see also Ref.~\cite{plebanski2006introduction}) is the general spherically symmetric solution of the Einstein equations with dust fluid $T_{ab}=\rho u_a u_b$
with $\rho$ and $u_a$ being the energy density and the four-velocity of the dust fluid. 
By using the geodesic slice and comoving radial coordinate $\tilde r$, we can write the spacetime metric as
\begin{align}\label{eq:LTBmetric}
    ds^2 &= - dt^2 + \frac{(\partial_{\tilde r}R)^2(t,\tilde{r})}{1+2E(\tilde{r})}d\tilde{r}^2 + R^2(t,\tilde{r}) d\Omega^2, 
\end{align}
where $E(\tilde{r})$ represents an arbitrary integral function restricted to $1+ 2E(\tilde{r})\geq 0$ to keep Lorentzian geometry.
The case of $1+ 2E= 0$ is allowed if $\partial_{\tilde r}R=0$ at the same radius $\tilde r$, sometimes called a ``bottle neck'' or ``wormhole''~\cite{plebanski2006introduction} which can make the areal radius $R$ non-monotonic function for the radius and can connect regions with the positive and negative root $\pm\sqrt{1+2E}$~\cite{Misner:1973prb,10.1143/PTP.72.1137, Escriva:2025eqc}.
The Einstein equations through the LTB metric are given by
\begin{align}
    (\partial_t R)^2(t,\tilde{r}) &= \frac{2M(\tilde{r})} {R(t,\tilde{r})} +2E(\tilde{r}),\label{eq:HamConst}\\ 
    \rho(t,\tilde{r}) &= \frac{1}{4\pi}\frac{\partial_{\tilde r}M(\tilde{r})}{R^2(t,\tilde{r}) \partial_{\tilde r}R(t,\tilde{r})},\label{eq:rho}
\end{align}
where $M(\tilde{r})$ is the second arbitrary integral function equivalent to the Misner--Sharp mass.\footnote{
From the Newtonian analogy with Eq.~\eqref{eq:HamConst}, $E$ and $M$ are sometimes called energy and mass functions, respectively, as the energy conservation law for a fluid element on the comoving radius $\tilde r$ (see, e.g., Ref.~\cite{galaxies10060112})
\begin{align*}
    \frac{1}{2}(\partial_t R)^2 - \frac{M}{R} = E.
\end{align*}
}
\footnote{The energy density remains non-singular for the case of $\partial_{\tilde r}R(t_s,\tilde r_s) =0= \partial_{\tilde r}M(\tilde r_s)= \partial_{\tilde r}E(\tilde r_s)$~\cite{Hellaby:1985zz}.
This situation of no singularity can be realized as long as we use non-pathological coordinates and restrict the shape of the arbitrary function $E(\tilde r)$, as choosing suitable initial data, to be $1+2E(r_s)=0$ so as a non-singular metric component in Eq.~\eqref{eq:LTBmetric}.
}
Here, we do not completely specify the radial coordinate $\tilde r$, and there is a gauge degree of freedom in choosing a new radial coordinate as a function of $\tilde r$ with the metric form and equations of motion unchanged.
One of $E(\tilde r)$ and $M(\tilde r)$ corresponds to the gauge degree of freedom.

In the long wavelength limit, identifying the radial coordinate $\tilde r$ with the isotropic coordinate $r$ in the initial time slice, we can write the metric as
\begin{align}
    ds^2 &\approx -dt^2 + \frac{(\partial_r R)^2(t,r)}{1+2E(r)}dr^2 + R^2(t,r) d\Omega^2. 
    \label{eq:E(r)inMetric}
\end{align}
Comparing Eqs.~\eqref{eq:zetaInMetric} and~\eqref{eq:E(r)inMetric}, we find the following relationship between  $E(r)$ and $\zeta(r)$ as
\begin{align}
     R &\approx a e^{\zeta}r,\\
    E &\approx \frac{1}{2}\qty(-1 + (1 + r \partial_r\zeta)^2).\label{eq:EfromZeta} 
\end{align}
This relation lets us derive the LTB function $E(r)$ directly from the curvature perturbation $\zeta(r)$ in the long-wavelength limit.
It should be noted that this relationship is valid only in the limit. 
As an example, we will plot $E(r)$ for the case of the profile $\zeta(r)=\mu\exp(-k^2r^2/2)$ in Fig.~\ref{fig:E(r)profile} in Sec.~\ref{sec:AnalyDemoWithExampleProfile}.

This function $M(r)$ can be expressed as
\begin{align}
    M &
    = \frac{R}{2}\qty((\partial_t R)^2 -2E)\\
    &= \frac{R}{2}\qty[\qty(H + 2\,\partial_t\ln\psi)^2R^2 -2E],
\end{align}
where 
the definition $R=a(t)\psi^2(t,r) r$ is used.
In the long-wavelength limit, $\zeta(r)=2\ln \psi$, then
\begin{align}
    M &\approx \frac{R}{2}\qty(H_i^2 R^2 - 2E)\\
    &\approx \frac{1}{2}a_i e^\zeta r\qty(H_i^2 a_i^2 e^{2\zeta} r^2 +1 - (1+r\partial_r\zeta)^2), \label{eq:MfromZeta}
\end{align}
where we have introduced the initial scale factor $a_i$ and Hubble parameter $H_i$ at an initial time $t=t_i$.
Then, the form of $M(r)$ can also be determined by the curvature perturbation $\zeta(r)$ as a function of the isotropic coordinate $r$.
This implies that the gauge degree of freedom for the radial coordinate $\tilde r$ has been fixed when we identify $\tilde r$ with $r$.

\subsection{Time evolution and future and past trapping horizons in the LTB spacetime}\label{subsec:EvolutionInLTB}
In this paper, we focus on gravitationally bound regions, characterized by $E(r)<0$, where gravitational instability can lead to PBH formation. 
For $E(r)<0$, the solution is known to be described as
\begin{empheq}[left=]{align}
    R(t,r) &= \frac{M(r)}{-2E(r)}\qty(1-\cos \eta), \\
    t &= t_\text{B}(r) + \frac{M(r)}{(-2E(r))^{3/2}}(\eta -\sin \eta),
\end{empheq}
where $\eta \in [0,2\pi]$ is a parameter and $t_B(r)$ is the third arbitrary integral function.
This parametric solution can be rewritten as
\begin{eqnarray}
    t&=&t_{\rm B}(r)+\frac{M(r)}{(-2E(r))^{3/2}}\cr
    &&\times\left\{
    \begin{array}{lcl}
     \cos^{-1}\qty(1+\frac{2E(r)R(t,r)}{M(r)}) -\sqrt{1-\qty(1+\frac{2E(r)R(t,r)}{M(r)})^2}&{\rm for}&t\leq t_{\rm max},      \\
     \pi+ \cos^{-1}\qty(-1-\frac{2E(r)R(t,r)}{M(r)}) +\sqrt{1-\qty(1+\frac{2E(r)R(t,r)}{M(r)})^2}&{\rm for}&t> t_{\rm max},\label{eq:t_pm}
    \end{array}
    \right.
\end{eqnarray}
where $t_{\rm max}$ is the time of the maximum expansion ($\eta=\pi$) given by $t_{\rm max}=t_{\rm B}(r)+\frac{\pi M}{\left(-2E\right)^{3/2}}$.%
\footnote{
    In other words, each expanding or collapsing phase follows the positive or negative root of Eq.~\eqref{eq:HamConst} for $\dot{R}$.
}
The $r$-dependent $t_B(r)$ involves decaying modes \cite{1977A&A....59...53S} (see also Ref.~\cite{plebanski2006introduction}), while the constant $t_B(r)$ excludes those, where the constant value can be eliminated through the constant shift of the time coordinate. Since we are interested in growing modes, we set $t_B(r)=0$ hereafter. 

Let us consider horizon trajectories in this spacetime in terms of trapping horizons defined by $\theta \theta'=0$ with $\theta$ and $\theta'$ being null expansions for two independent radial-future null directions. 
Following Ref.~\cite{Uehara:2024yyp}, we classify the trapping horizons into past, bifurcating, and future trapping horizons for $\theta + \theta' > 0$, $=0$, and $<0$, respectively.

In the LTB spacetime, we have future and past trapping horizons, whose trajectories can be described by $t=t_{\rm TH+}(r)$ and $t=t_{\rm TH-}(r)$, respectively. 
In a spherically symmetric spacetime, it can be shown that the radius of a trapping horizon is given by twice the Misner-Sharp mass~\cite{PhysRevD.49.6467,galaxies10060112}, $R=2M$. 
Therefore we obtain
\begin{empheq}[left={t =}\empheqlbrace]{align}
    &t_{\rm TH-}(r) \coloneqq \frac{M(r)}{(-2E(r))^{3/2}}\left[\cos^{-1}\qty(1+4E(r)) -\sqrt{1-\qty(1+4E(r))^2}\right], \label{eq:AH-}\\
    &t_{\rm TH+}(r)\coloneqq\frac{M(r)}{(-2E(r))^{3/2}}\left[\pi+ \cos^{-1}\qty(-1-4E(r)) +\sqrt{1-\qty(1+4E(r))^2}\right].\label{eq:AH+}
\end{empheq}
When we choose profiles of $E(r)$ and $M(r)$ as initial data, the evolution of this system and also the past and future trapping horizons $t=t_{\rm TH-}(r)$ and $t=t_{\rm TH+}(r)$ are determined. 
In terms of the LTB solution $t=t_\pm(r)$ as Eqs.~\eqref{eq:t_pm}, we define \emph{type B} structure of spacetime by the existence of a bifurcating trapping horizon $r=r_b$ where the past and future trapping horizons $t=t_{\rm TH\pm}(r_b)$ meet as
\begin{align}
    t_{\rm TH+}(r_b)-t_{\rm TH-}(r_b) = 0,
\end{align}
while the past and future trapping horizons do not meet for \emph{type A}.

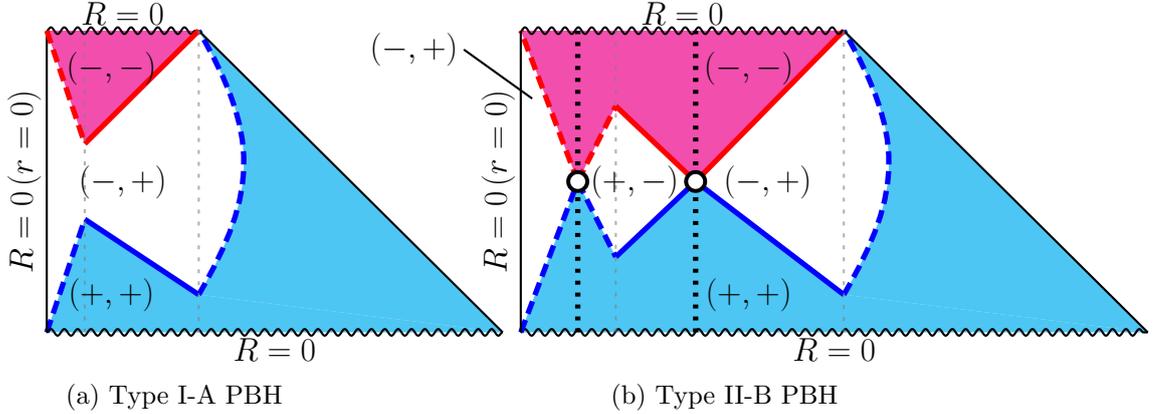
\begin{figure}[htbp]
  \centering
    \begin{minipage}{0.3\linewidth}
        \centering
        \scalebox{.5}{
            \begin{tikzpicture}[ultra thick]
                \coordinate(i0)at(10,-4);
                \coordinate(i+)at(2,4);
                \coordinate(i-)at(2,-4);
              
                \coordinate(i++)at(-2,4);
                \coordinate(i--)at(-2,-4);
                \coordinate(LEH)at(-1,1);
                \coordinate(LBEH)at(-1,-1);

                \fill[cyan,opacity=.7](i--)--(LBEH)--(2,-3)--(i0)--cycle;
                \fill[cyan,opacity=.7,bend right,distance=90pt](2,-3)to(i+)--(i0)--cycle;
                \draw[blue,dash pattern=on 10pt off 5pt,line width=4pt,bend right,distance=90pt](2,-3)to(i+);
                \draw[blue,dash pattern=on 10pt off 5pt,line width=4pt](i--)--(LBEH);
                \draw[blue,line width=4pt](LBEH)--(2,-3);
                
                \draw[snake it](i--)--(i0);
                \path(i--)--node[below]{\Huge $R=0$}(i0);
                
                \path(i--)--node[sloped, above]{\Huge $R=0 \,
                (r=0)$}(i++);
              
                \fill[magenta,opacity=.7](i++)--(LEH)--(i+)--cycle;
                
                \draw[snake it](i++)--(i+);
                \path(i++)--node[above]{\Huge $R=0$}(i+);
                \path(i++)--(i--);
                
                \draw(i++)--(i--);
                \draw(i0)--(i+);
                \draw[red,dash pattern=on 10pt off 5pt,line width=4pt](i++)--(LEH);
                \draw[red,line width=4pt](LEH)--(i+);

                \draw[loosely dashed,gray, opacity=.6](-1,-4)to(-1,4);
                \draw[loosely dashed, gray, opacity=.6](i+)--(i-);

              
                \path[magenta](-.9,2)--node[above]{}(0,0);
                \path[magenta](-.9,-4.5)--node[above]{}(0,0); 
              
                \path(-0.3,3)node[]{\Huge $(-,-)$};
                \path(0,0)node[black]{\Huge $(-,+)$};
                \path(-0.3,-3)node[]{\Huge $(+,+)$};
              \end{tikzpicture}
        }
        \subcaption{Type I-A PBH}
    \end{minipage}
    \begin{minipage}{0.62\linewidth}
        \centering 
        \scalebox{.5}{
        \begin{tikzpicture}[ultra thick]
          \coordinate(i0)at(13,-4);
          \coordinate(i+)at(5,4);
          \coordinate(i-)at(5,-4);
        
          \coordinate(Sch_flat)at(5,-3);
        
          \coordinate(i++)at(-3.5,4);
          \coordinate(i--)at(-3.5,-4);
        
          \coordinate(LEH)at(-1,-2);
        
          \coordinate(LBEH)at(-1,2);
        
          \coordinate(bifurL)at(-2,0);
          \coordinate(bifurR)at(1.1,0);
        
          \fill[cyan,opacity=.7](i--)--(bifurL)--(LEH)--(bifurR)--(Sch_flat)--(i0)--cycle;
          \fill[cyan,opacity=.7,bend right,distance=90pt](Sch_flat)to(i+)--(i0)--cycle;
          \draw[blue,dash pattern=on 10pt off 5pt,line width=4pt,bend right,distance=90pt](Sch_flat)to(i+);

          \draw[snake it](i--)--(i0);
          \path(i--)--node[below]{\Huge $R=0$}(i0);
          \path(i--)--node[sloped, above]{\Huge $R=0 \,
          (r=0)$}(i++);
        
          \fill[magenta,opacity=.7](i++)--(bifurL)--(LBEH)--(bifurR)--(i+);
          \draw[red,line width=4pt](bifurR)--(i+);
          \draw[red,dash pattern=on 10pt off 5pt,line width=4pt](bifurL)--(i++);
          \draw[blue,dash pattern=on 10pt off 5pt,line width=4pt](bifurL)--(LEH);
          \draw[blue,line width=4pt](LEH)--(bifurR);
          \draw[blue,line width=4pt](bifurR)--(Sch_flat);
          \draw[red,dash pattern=on 10pt off 5pt,line width=4pt](bifurL)--(LBEH);
          \draw[red,line width=4pt](LBEH)--(bifurR);
          \draw[blue,dash pattern=on 10pt off 5pt,line width=4pt](i--)--(bifurL);
        
          \draw[snake it](i++)--(i+);
          \path(i++)--node[above]{\Huge $R=0$}(i+);
          
          \draw(i++)--(i--);
          \draw(i+)--(i0);
          \draw[loosely dashed,gray, opacity=.6](-1,-4)to(-1,4);
        
          \draw[loosely dashed, gray, opacity=.6](i+)--(i-);
        
          \draw[line width=4pt, loosely dashed](-2,-4)--(-2,4);
          \draw[line width=4pt, loosely dashed](1.1,-4)--(1.1,4);
          \filldraw[fill=white,line width=3pt](bifurL)circle[radius=0.25];
          \filldraw[fill=white,line width=3pt](bifurR)circle[radius=0.25];
        
          \path(2.5,3)node[]{\Huge $(-,-)$};
        \draw(-3.2,2.2)--(-5,3.5)node[left]{\Huge $(-,+)$};
          \path(3,0)node[black]{\Huge $(-,+)$};
          \path(-.5,0)node[black]{\Huge $(+,-)$};
          \path(2.5,-3)node[]{\Huge $(+,+)$};
        \end{tikzpicture}
        }
        \subcaption{Type II-B PBH}
    \end{minipage}
  \caption{
        Conformal diagrams of PBH for dust case as a patchwork of the three diagrams (closed FLRW, Schwarzschild, and flat FLRW) and trapping horizons.
        The left panel defines a type I-A PBH formation.
        The right one is that for a type II-B PBH.
        The combination of signs of the null expansion $(\theta,\theta')$ separates each region.
        Red- and blue-shaded regions represent future ($-,-$) and past ($+,+$) trapped regions, respectively. 
        Future-outer, past-outer, future-inner, and past-inner trapping horizons are denoted by red-solid, blue-solid, red-dashed, and blue-dashed lines, respectively.
        The bifurcating
        trapping horizons $(\theta = \theta' = 0)$ are depicted as circles where the future and past trapping horizons meet, also locating on the 
        black dashed lines of stationary point of $R$ for $r$, $\partial_r R(t,r)=0$.
    }
  \label{fig:PBHponchi}
\end{figure}

As an example, in Fig.~\ref{fig:PBHponchi}, let us show the conformal diagrams of PBH formation for the dust case given by a patchwork of closed FLRW, Schwarzschild, and flat FLRW regions in order from the inside. 
    All of them are subclasses of the LTB solution with each specific shape of $E(r)$ and $M(r)$.
The vertical gray dashed lines separate the patched regions from each other. 
The thick red and blue lines are future and past trapping horizons, respectively.
    The black dashed lines represent trajectories of the ``neck'' or ``belly'' of $\partial_r R(t,r)=0$ and $1+2E(r)=0$.
The left and right panels show type I and II primordial fluctuations, respectively. 
One can find bifurcating trapping horizons depicted as circles for type II fluctuation but not for type I. 
In the next section, we will analytically show that this behavior is general for the growing mode LTB solution, regardless of the fluctuation profile. 

This kind of configuration is similar to the notion of `semiclosed world,' which is constructed by patching a part of the closed FLRW and of the maximally extended Schwarzschild solution, including the bifurcation two-sphere (e.g. Ref.~\cite{BA67359925}); unlike the usual formation model, the maximum expansion and the contraction process in the closed FLRW region cannot be observed by a distant observer because the event horizon covers the sequences~\cite{Zeldovich:1983cr,Novikov:1989sz,Frolov:1998wf,1963JETP...16..732Z,1964SvA.....7..587N} (see also Ref.~\cite{gourgoulhonBH}).

\section{The equivalence of type I/II and A/B for the LTB solution}\label{sec:NoGap}
For some cases of type II fluctuations, the resultant spacetime has bifurcating trapping horizons, where past and future 
trapping
horizons meet~\cite{PhysRevD.83.124025,Uehara:2024yyp}. 
This is called the type B configuration of 
trapping 
horizons for a resultant spacetime, not for the initial perturbation type (type I/II). 
Therefore, we can summarize each case using the combination of initial fluctuation type (I/II) and resultant horizon configuration (A/B), such as type II-A and II-B.
In our previous work~\cite{Uehara:2024yyp} for a radiation fluid system with the initial spatial profile of curvature perturbation $\zeta(r)=\mu\exp(-k^2r^2/2)$, we numerically found that the cases with the amplitude $\mu\geq \mu_{II}=e/2 = 1.359\cdots$ correspond to type II and the cases with $\mu >\mu_{B}\sim 1.8$ cause type B through numerical relativity simulations of the non-linear evolution.
The threshold $\mu_B$ generally depends on the initial profile $\zeta(r)$ and on the equation of state for the matter content, as KHW had anticipated, whereas $\mu_{II}$ only depends on the initial profile $\zeta(r)$. 
In the following subsections, however, we show that 
$\mu_{II}=\mu_{B}$, so that type I/II and A/B classifications are equivalent to each other for the LTB solution, regardless of the functional forms of $M(r)$ and $E(r)$.

Let a non-singular coordinate system cover a region in a spherically symmetric dust solution for the case of $E<0$, and the initial data is given as a long wavelength solution sufficiently near the Big Bang time.
Then, the type B structure of trapping horizon configuration, bifurcating trapping horizon, forms if and only if it results from the initial data with a stationary point of $R(r)$ as a type II fluctuation.
This is shown in the following subsections.

\subsection{A stationary point implies a bifurcating trapping horizon}\label{subsec:SuffCond}
In this subsection, we show that the existence of a stationary point (type II fluctuation) is a sufficient condition for the existence of a bifurcating trapping horizon (type B configuration) following Ref.~\cite{PhysRevD.69.043502} (or the textbook of Krasinski--Plebanski~\cite{plebanski2006introduction}). 
In other words, we show that the existence of stationary points $r=r_{p}$ of $R(r)$ with $\partial_rR(r_p) \approx 0$ leads to the meeting of past and future trapping horizons $t=t_{\rm TH-}(r_p)$ and $t=t_{\rm TH+}(r_p)$ as $t_{\rm TH+}(r_p)-t_{\rm TH-}(r_p)=0$. 

At $r=r_p$, we obtain $E(r_p) \approx -1/2$ from Eq.~\eqref{eq:EfromZeta} and substitute this into the horizon trajectories Eq.~\eqref{eq:AH-} and \eqref{eq:AH+}, then
\begin{empheq}[left=]{align}
    t_{\rm TH-}(r_p) &= M(r_p)\cos^{-1}\qty(-1)=\pi M(r_p) ,\label{eq:tAH-AtNeck} \\ 
    t_{\rm TH+}(r_p) &= M(r_p) \qty(\pi+\cos^{-1}\qty(+1) ) =\pi M(r_p)=t_{\rm TH-}(r_p).\label{eq:tAH+AtNeck}
\end{empheq}
This indicates that the past and future trapping
horizons $t_{\rm TH-}(r)$ and $t_{\rm TH+}(r)$ converge at $(t,r)=(t_{\rm TH-}(r_p),r_p)=(t_{\rm TH+}(r_p),r_p)$.
The specific example will be shown in the next section. 
This result confirms that type II fluctuations lead to type B PBHs, characterized by bifurcating trapping horizons. 
That is, type II-A PBH formation cannot be realized for dust cases differently from the radiation-dominated cases reported in Ref.~\cite{Uehara:2024yyp}.

\subsection{A bifurcating trapping horizon implies a stationary point}\label{subsec:curvatureFluc}
Conversely, we now establish that the stationary point is necessary for the bifurcating trapping horizons. 
Assuming the type B horizon structure, we focus on a bifurcating trapping horizon at the radius $r=r_b$. 
By definition, at $r=r_b$, the past and future trapping horizons match ($t_{\rm TH-}(r_b) = t_{\rm TH+}(r_b)$). 
Then, using Eqs.~\eqref{eq:AH-} and \eqref{eq:AH+}, we obtain  
\begin{align}
    0 &= t_{\rm TH+}(r_b) - t_{\rm TH-}(r_b),\cr
    &=\frac{M(r_b)}{(-2E(r_b))^{3/2}} \qty[\pi - \cos^{-1}(1+4E(r_b)) +\cos^{-1}(-(1+4E(r_b)))+2\sqrt{1- \qty(1 +4E(r_b))^2 }],\cr
    &\eqqcolon \frac{M(r_b)}{(-2E(r_b))^{3/2}} y(x),\label{eq:necEq}
\end{align}
where in the last line we have put $x=1+4E(r_b)$ and $y(x) \coloneqq \pi - \cos^{-1}x + \cos^{-1}(-x) + 2\sqrt{1-x^2}$.
Taking the derivative of $y(x)$, we obtain
\begin{align}
    y'(x)=2\sqrt{\frac{1-x}{1+x}}.
\end{align}
We find the function $y=y(x)$ is continuous for $-1\le x\le 1$ and monotonically increasing because $y'(x) >0$ for $-1< x<1$.
So $y(x)\ge y(-1)=\pi-\pi+0=0$.
Then, $x=-1$ or $E(r_b)=-1/2$ is uniquely obtained from the condition of a bifurcating trapping horizon as Eq.~\eqref{eq:necEq}.

In the context of PBH formation, considering the adiabatic growing mode, the curvature fluctuation $\zeta$ in the long-wavelength limit corresponds to the function $E(r)$.
Then, the relation $E(r_b)=-1/2$ leads to $(r \partial_r\zeta)|_{r= r_b} = -1$ from Eq.~\eqref{eq:EfromZeta}.
Since $\partial_r R=\partial_r (ar \exp \zeta)=a\partial_r (1+r\partial_r\zeta)\exp\zeta$, we can conclude that $\partial_r R(r_b)=0$. 
This analysis, together with the result of the preceding subsection, establishes the equivalence of the classifications type I/II and type A/B in the dust case. 
This is consistent with the findings of KHW~\cite{PhysRevD.83.124025}.

\section{Demonstration for an example profile of a curvature perturbation}\label{sec:AnalyDemoWithExampleProfile}
\begin{figure}[htbp]
    \centering
    \includegraphics[width=.49\linewidth]{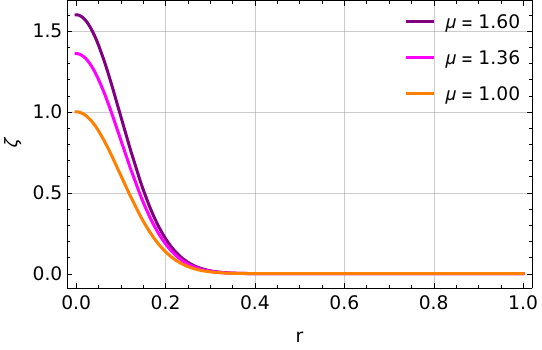}
    \includegraphics[width=.49\linewidth]{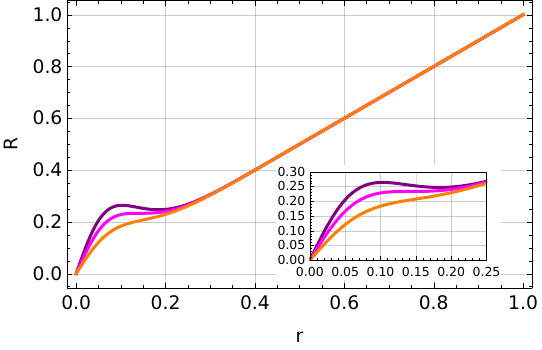}
    \caption{
        Spatial profiles of curvature perturbation $\zeta(r) = \mu \exp(-k^2 r^2/2)$ on the left panel and areal radius $R = ae^{\zeta(r)} r$ on the right panel for different values of $\mu$, where we have set $k=10$. 
        The areal radius $R$ 
        has a stationary point
        for $\mu \ge e/2=1.359\cdots$.
    }
    \label{fig:AzetaAndRprofile}
\end{figure}

To demonstrate the theoretical results, we consider a specific curvature fluctuation profile: 
\begin{align}
    \zeta(r) = \mu e^{-\frac{1}{2}k^2r^2}, \label{eq:expprofile}
\end{align}
where $k$ represents a characteristic wave number.
The initial profile and corresponding areal radius are shown in Fig.~\ref{fig:AzetaAndRprofile}.
The left panel illustrates the spatial profile of the curvature perturbation $\zeta(r)$ for different amplitude values $\mu$.
The right panel shows the corresponding areal radius $R=ar\exp\zeta(r)$, which has a stationary point for $\mu \ge e/2$ and is classified as type II. 

\subsection{Analytical example: the LTB solution}
\begin{figure}[htbp]
    \centering
    \includegraphics[width=.49\linewidth]{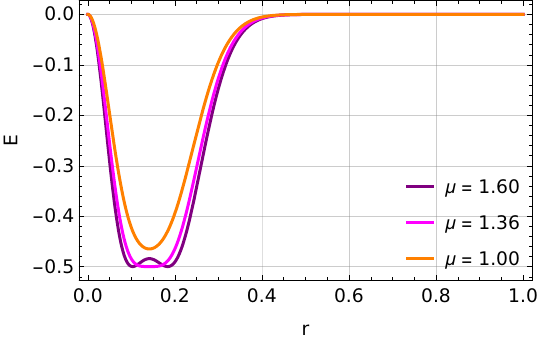}
\includegraphics[width=.49\linewidth]{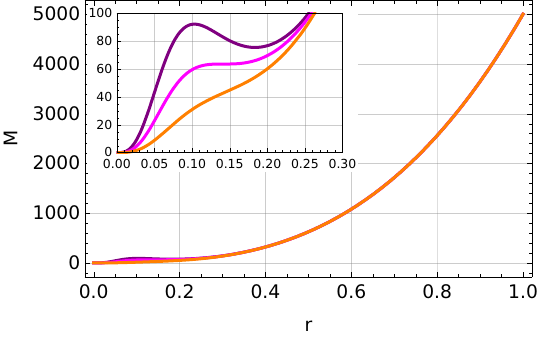}
    
    \caption{
        Profiles of the functions $E(r)$ and $M(r)$ in the LTB system which is determined from  $\zeta(r) = \mu\exp(-k^2r^2/2)$ with $k=10$.
    }
    \label{fig:E(r)profile}
\end{figure}

Using the curvature perturbation profile \eqref{eq:expprofile}, we calculate the LTB functions $E(r)$ and $M(r)$ through Eqs.\eqref{eq:EfromZeta} and \eqref{eq:MfromZeta}.
These profiles are plotted in Fig.~\ref{fig:E(r)profile} for different values of $\mu$.
The zoomed-in view highlights the neck and belly as local minimum and maximum for $\mu>e/2=1.359\cdots$.
One of the plot curves is for $\mu=1.36$, which is of type II and close to the marginal case with a stationary point but not an extreme point.

\begin{figure}[htbp]
    \centering
    \includegraphics[width=.7\linewidth]{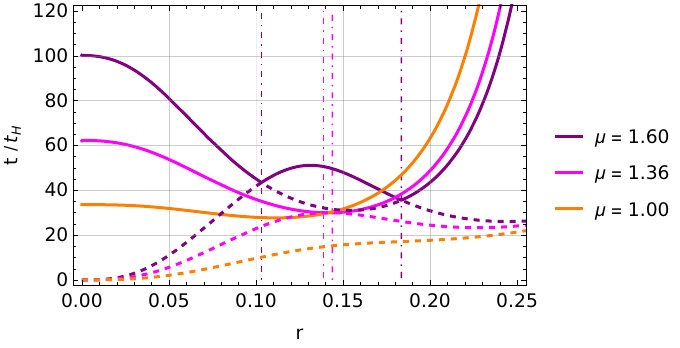}
    
    \caption{
        Trajectories of trapping horizons $t = t_{\rm TH\pm}(r)$ in the
        LTB solution specified by the functional form of $\zeta(r)$ given by Eq.~\eqref{eq:expprofile}. 
        The solid and dashed curves correspond to future and past trapping horizons. 
        The vertical dot-dashed lines describe the trajectories of stationary points of $R$ in comoving radial coordinate $r$ in the synchronous comoving coordinates for the LTB solution for $\mu=1.36$ and $1.60$, respectively.
        The case of $\mu=1.00$ (type I) has no stationary point.
        The vertical axis is normalized by horizon entry time $t=t_H$ defined by $k=a(t_H)H(t_H)$. 
    }
    \label{fig:AHltb}
\end{figure}

Fig.~\ref{fig:AHltb} shows the trajectories of the trapping horizons derived from the initial curvature perturbation profile. 
The solid and dashed curves represent the future and past trapping horizons. 
The vertical dot-dashed lines describe the trajectories of the stationary points of $R$.
Here, $t=t_H$ is the horizon entry time, defined as $k=a(t_H)H(t_H)$ (see, e.g., Appendix in Ref.~\cite{Yoo:2024lhp} for the explicit expression of $t_H$).

\subsection{Numerical example: Trajectories of stationary point for different equations of states}\label{sec:NeckDemo}
In this subsection, we illustrate the influence of the equation of state on horizon dynamics and PBH classification, demonstrating the evolution of stationary points trajectories for different equations of state given by $p=w\rho$ characterized by the linear coefficient $w$ with $p$ and $\rho$ being the pressure and energy density, respectively. 
We perform the numerical simulation in a similar way to that in our previous work~\cite{Uehara:2024yyp}, changing the equation of state parameter $w$\footnote{
    We use the numerical code COSMOS-S publicly available at \texttt{https://sites.google.com/view/cosmoscode/}~\cite{PhysRevD.105.103538,PhysRevLett.111.161102,PhysRevD.89.104032}
}.

\begin{figure}[htbp]
    \centering
    \begin{minipage}{0.43\linewidth}
        \centering
        \includegraphics[width=\linewidth]{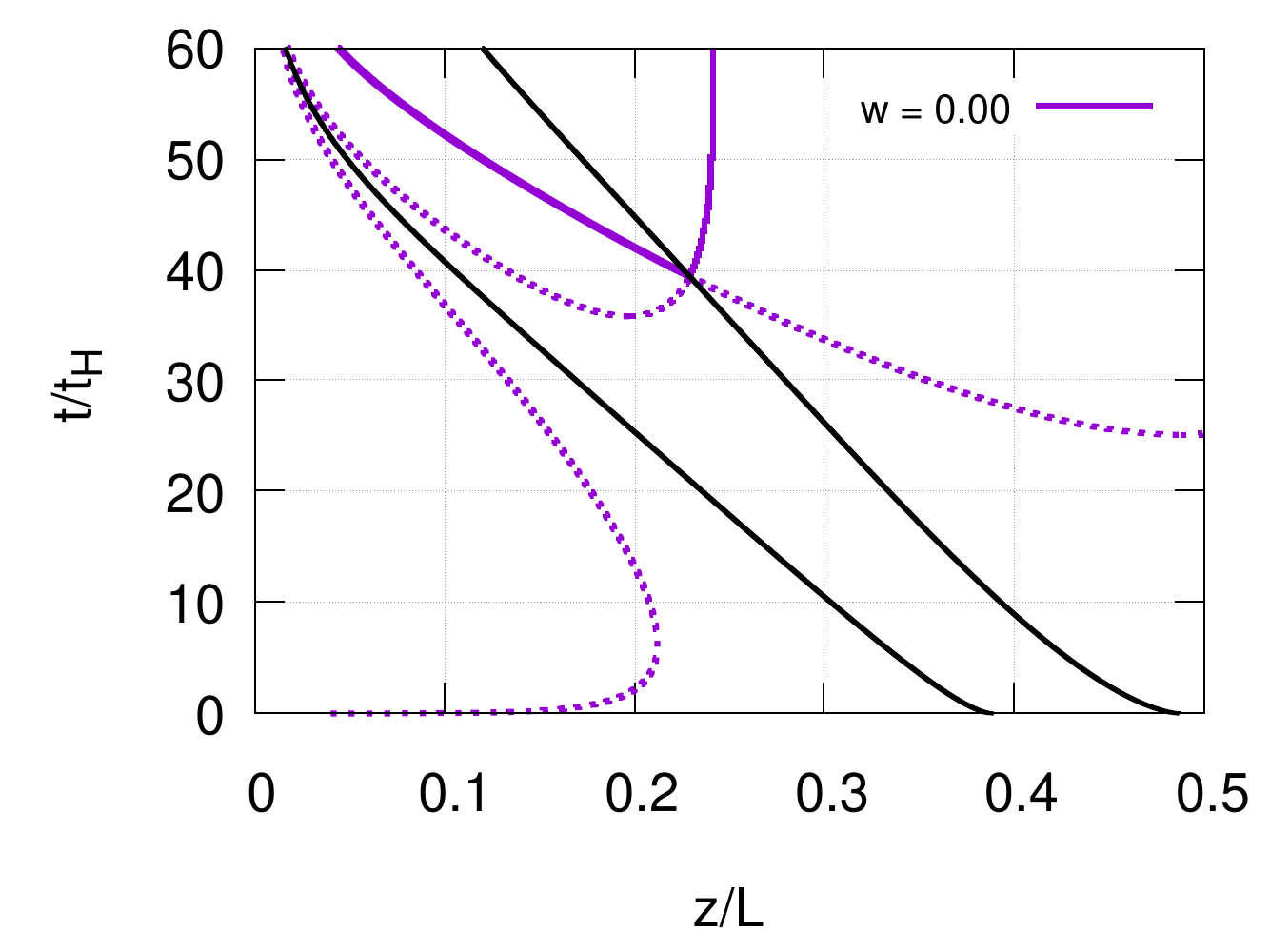}
        \subcaption{$w=0.00$}
        \label{fig:softerEoS00}
    \end{minipage}
    \begin{minipage}{0.43\linewidth}
        \centering
        \includegraphics[width=\linewidth]{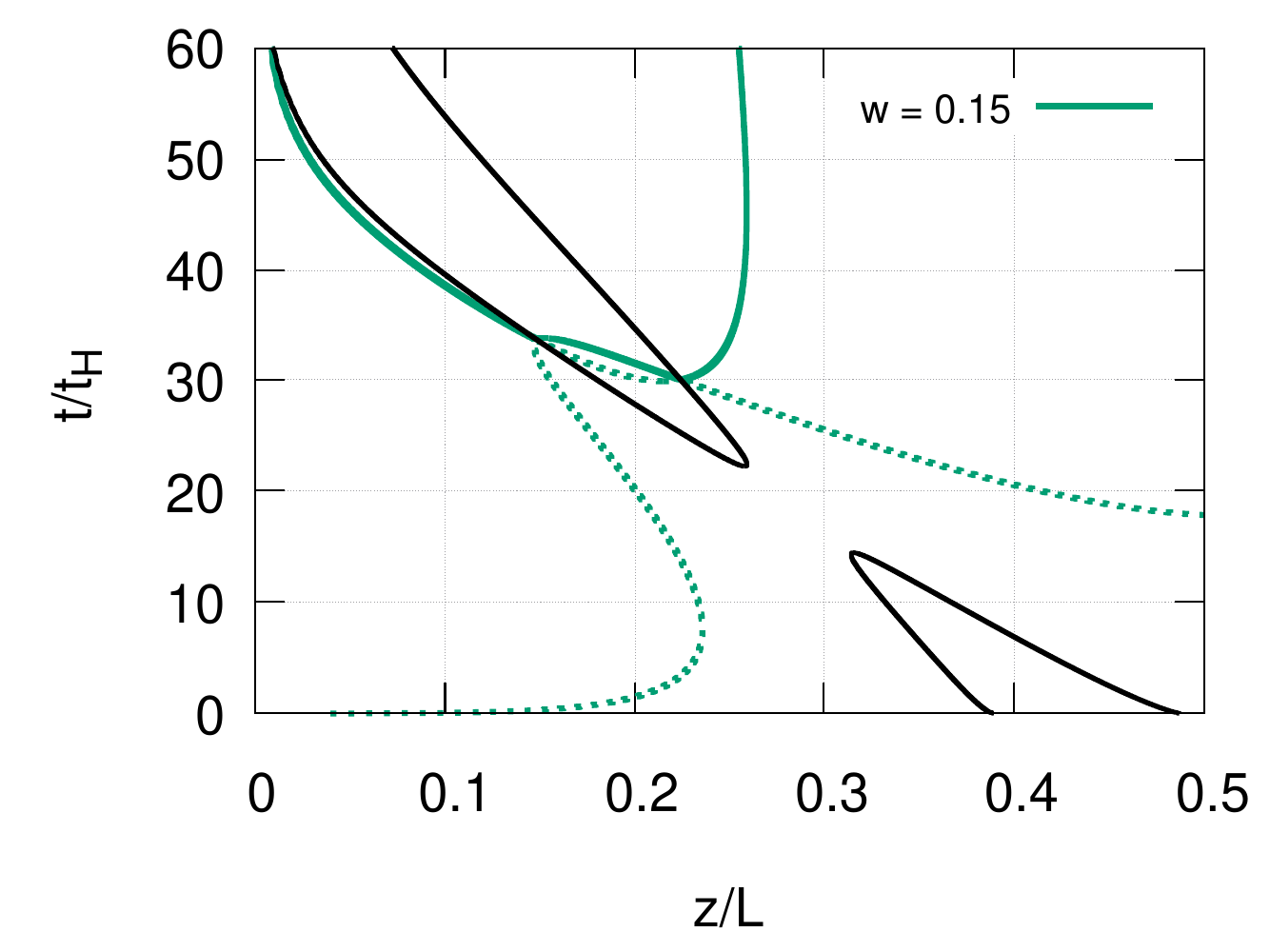}
        \subcaption{$w=0.15$}
        \label{fig:softerEoS015}
    \end{minipage}

    \begin{minipage}{0.43\linewidth}
        \centering
        \includegraphics[width=\linewidth]{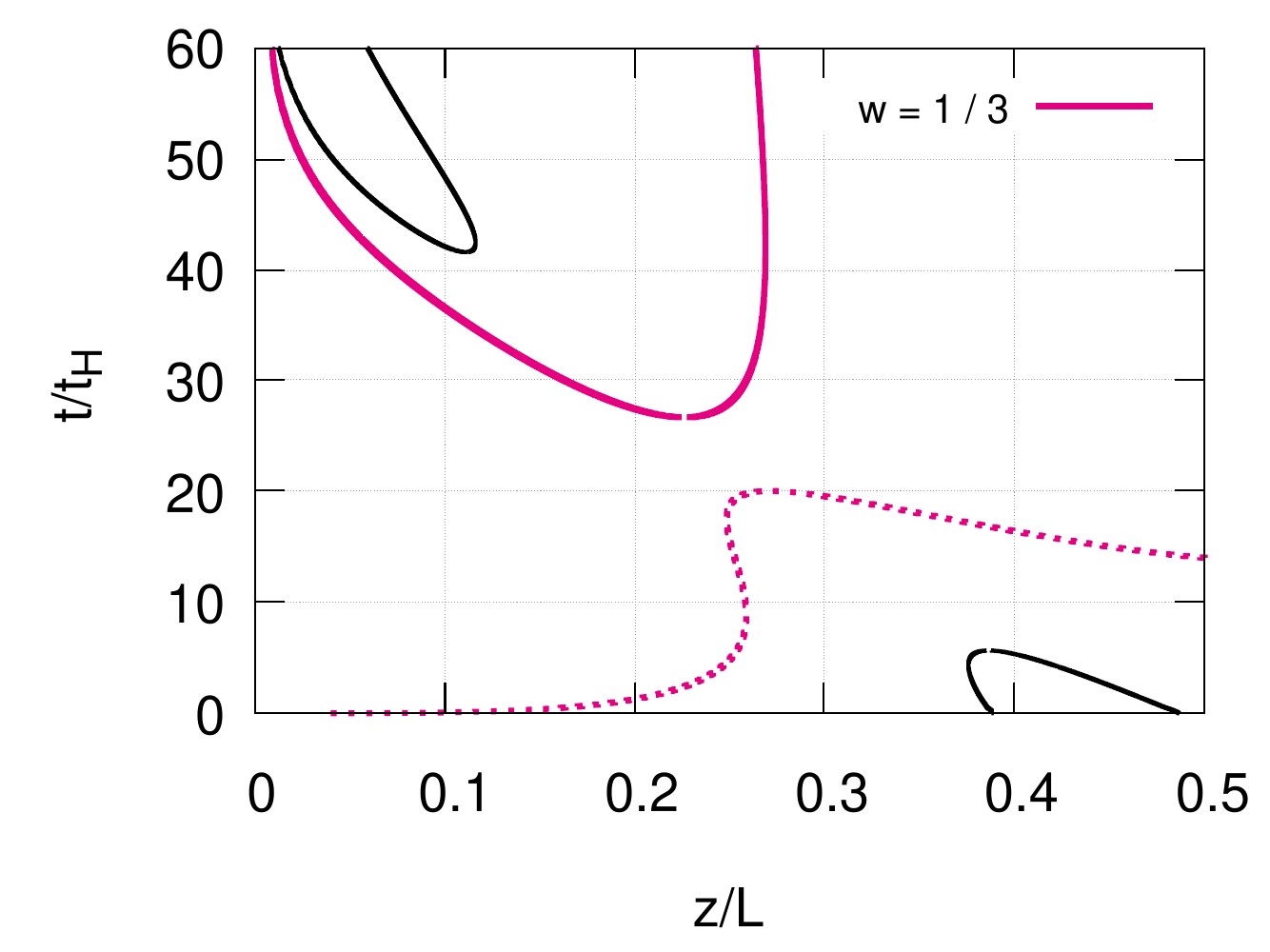}
        \subcaption{$w=1/3$}
        \label{fig:softerEoS13rd}
    \end{minipage}
    \begin{minipage}{0.43\linewidth}
        \centering
        \includegraphics[width=\linewidth]{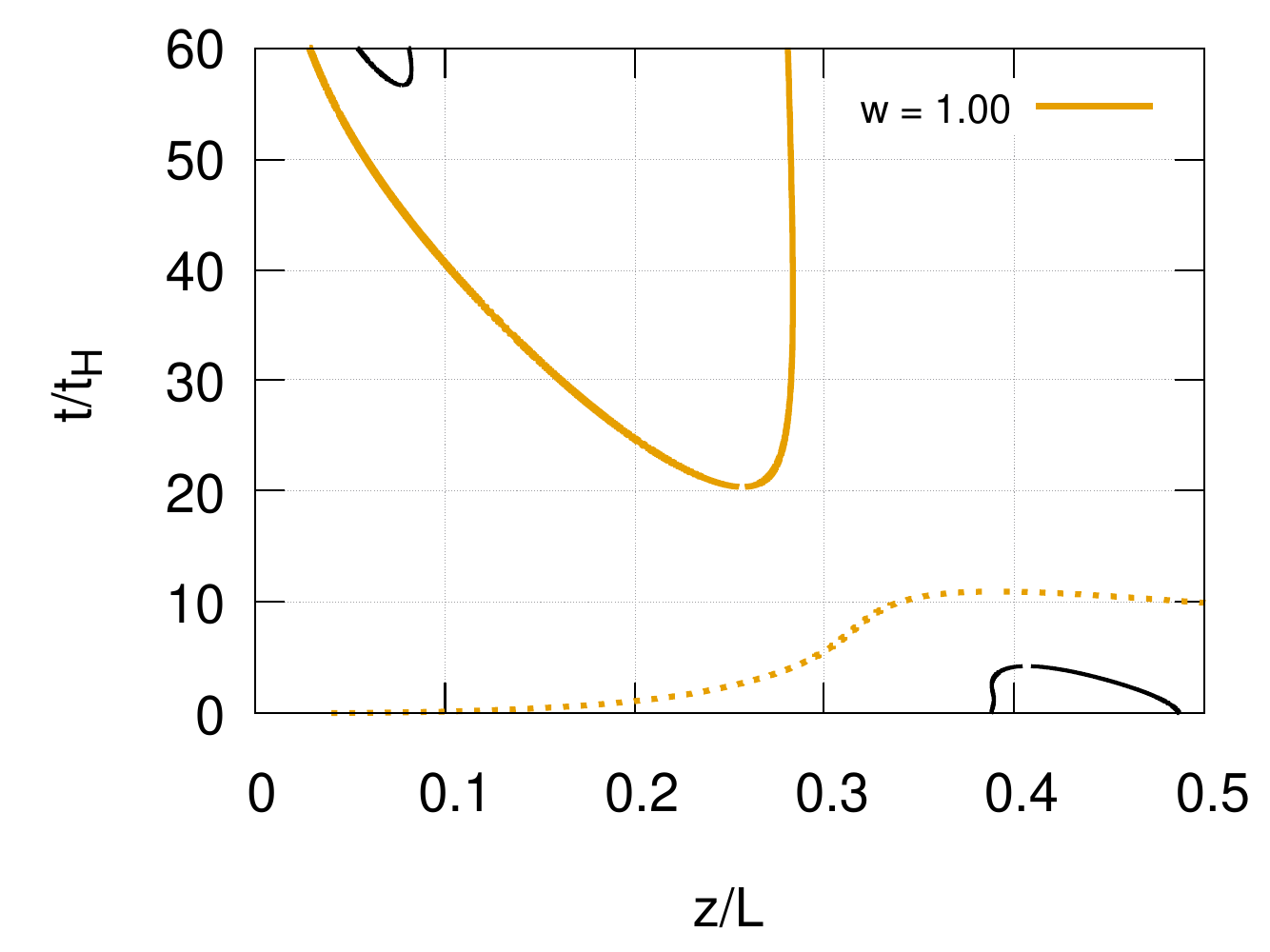}
        \subcaption{$w=1.00$}
        \label{fig:softerEoS100}
    \end{minipage}
    \caption{
     Numerical results of the trajectories of future/past trapping horizons (solid/dashed curves) and stationary points (black curves) for different values of $w$ with fixed $\mu=1.60$.
    The horizontal axis is the radial coordinate $z$, which rescales the radial coordinate $r$ through Eq.~(2.3) in Ref.~\cite{Uehara:2024yyp}. 
    $z$ is normalized by the initial size of the numerical box $L$ (see Ref.~\cite{Uehara:2024yyp} for details).
    }
    \label{fig:softerEoS}
\end{figure}

In Fig.~\ref{fig:softerEoS}, we show the horizon and stationary point trajectories for type II fluctuation with $\mu =1.60$ changing the equation of state parameter $w$ for different values in the range $w \in [0.0,1.0]$. 
For these cases, the amplitude $\mu$ is over the threshold for black hole formation. 
For the case of $w=0.00$ (Fig.~\ref{fig:softerEoS00}), as in the LTB solution in Fig.~\ref{fig:AHltb}, the stationary points of the neck and belly structure persist during the whole time evolution. 
The future and past trapping horizons contact on the trajectory of the stationary 
points, and the contact point corresponds to the bifurcating trapping horizon. 
For relatively stiffer 
$w$ (Fig.~\ref{fig:softerEoS015}), the trajectories of the stationary points 
disappear at some moment and then reappear after some finite time. 
For even stiffer $w$ (Figs.~\ref{fig:softerEoS13rd} and ~\ref{fig:softerEoS100}), large pressure gradients prevent the formation of a bifurcating trapping horizon. 
Thus, it changes from type B to type A. 
These results indicate that a stiffer $w$ increases the threshold $\mu_{B}$ for type B PBH formation, similar to the effect on the threshold $\mu_{A}$ for black hole formation \cite{Musco:2012au, Escriva:2020tak}.

\section{Conclusion}\label{sec:conclusion}
This work investigates the relationship between type I/II fluctuations and type A/B structures. 
We discussed how the stationary point in the initial spacelike hypersurface with high-amplitude primordial fluctuations in the long-wavelength regime leads to the formation of bifurcating trapping horizons in the dust case, using the LTB solution, and vice versa.

In the pioneering work \cite{PhysRevD.83.124025} on type I/II PBH formation, the authors did not distinguish type I/II and A/B. 
They presented two examples of type I-A and II-B PBH formation, characterized by a specific profile of spherical inhomogeneity. 
As is reported in Ref.~\cite{Uehara:2024yyp}, type II fluctuation does not necessarily result in type B PBH formation.

In this paper, we have proved that, for the dust case ($w=0$), type I/II and A/B classifications are equivalent regardless of the fluctuation profile. 
Our finding contrasts with the case where pressure exists with $w>0$.

This study builds on earlier work using the LTB model, highlighting the effects of the neck structure associated with the type II fluctuations on the PBH formation dynamics.
In addition, we clarify the significance of the pressure gradient from the change of the equation of state parameter $w$ on the PBH formation dynamics in the context of the type I/II and A/B classifications, showing that type B PBH formation is significantly suppressed in the
presence of a pressure gradient. 


\begin{acknowledgments}
The authors are grateful to Masaaki Shimada for useful discussions.
KU would like to take this opportunity to thank the “THERS Make New Standards Program for the Next Generation Researchers” supported by JST SPRING, Grant Number JPMJSP2125. 
AE thanks the support from the YLC program at the Institute for Advanced Research, Nagoya University.
This work was supported by JSPS KAKENHI Grant Numbers 20H05850~(CY), 20H05853~(TH, CY),  24KJ1223~ (DS), 24K07027~(TH, CY), and 25K07281~(CY). 
\end{acknowledgments}

\bibliographystyle{JHEP}
\bibliography{ref}

\providecommand{\href}[2]{#2}\begingroup\raggedright\begin{thebibliography}{10}

\bibitem{Oppenheimer:1939ue}
J.R.~Oppenheimer and H.~Snyder, \emph{{On Continued gravitational contraction}}, \href{https://doi.org/10.1103/PhysRev.56.455}{\emph{Phys. Rev.} {\bfseries 56} (1939) 455}.

\bibitem{Zeldovich:1967lct}
Y.B.~Zel'dovich and I.D.~Novikov, \emph{{The Hypothesis of Cores Retarded during Expansion and the Hot Cosmological Model}}, {\emph{Soviet Astron. AJ (Engl. Transl. ),} {\bfseries 10} (1967) 602}.

\bibitem{10.1093/mnras/152.1.75}
S.~Hawking, \emph{{Gravitationally Collapsed Objects of Very Low Mass}}, \href{https://doi.org/10.1093/mnras/152.1.75}{\emph{Monthly Notices of the Royal Astronomical Society} {\bfseries 152} (1971) 75} [\href{https://arxiv.org/abs/https://academic.oup.com/mnras/article-pdf/152/1/75/9360899/mnras152-0075.pdf}{{\ttfamily https://academic.oup.com/mnras/article-pdf/152/1/75/9360899/mnras152-0075.pdf}}].

\bibitem{10.1093/mnras/168.2.399}
B.J.~Carr and S.W.~Hawking, \emph{{Black Holes in the Early Universe}}, \href{https://doi.org/10.1093/mnras/168.2.399}{\emph{Monthly Notices of the Royal Astronomical Society} {\bfseries 168} (1974) 399} [\href{https://arxiv.org/abs/https://academic.oup.com/mnras/article-pdf/168/2/399/8079885/mnras168-0399.pdf}{{\ttfamily https://academic.oup.com/mnras/article-pdf/168/2/399/8079885/mnras168-0399.pdf}}].

\bibitem{Carr:1975qj}
B.J.~Carr, \emph{{The Primordial black hole mass spectrum}}, \href{https://doi.org/10.1086/153853}{\emph{Astrophys. J.} {\bfseries 201} (1975) 1}.

\bibitem{PhysRevD.71.104009}
T.~Harada and B.J.~Carr, \emph{Upper limits on the size of a primordial black hole}, \href{https://doi.org/10.1103/PhysRevD.71.104009}{\emph{Phys. Rev. D} {\bfseries 71} (2005) 104009}.

\bibitem{PhysRevD.83.124025}
M.~Kopp, S.~Hofmann and J.~Weller, \emph{Separate universes do not constrain primordial black hole formation}, \href{https://doi.org/10.1103/PhysRevD.83.124025}{\emph{Phys. Rev. D} {\bfseries 83} (2011) 124025}.

\bibitem{osti_4173295}
C.W.~Misner, \emph{Gravitational collapse.}, {\emph{pp 115-215 of Astrophysics and General Relativity. Vol. I. / Chretien, M. (ed.). New York Gordon and Breach, Science Publishers, Inc.} (1969) }.

\bibitem{A_Barnes_1970}
A.~Barnes, \emph{On gravitational collapse against a cosmological background}, \href{https://doi.org/10.1088/0305-4470/3/6/007}{\emph{Journal of Physics A: General Physics} {\bfseries 3} (1970) 653}.

\bibitem{10.1093/mnras/207.1.23P}
Y.B.~Zel'dovich and L.P.~Grishchuk, \emph{Structure and future of the ‘new’ universe}, \href{https://doi.org/10.1093/mnras/207.1.23P}{\emph{Monthly Notices of the Royal Astronomical Society} {\bfseries 207} (1984) 23P} [\href{https://arxiv.org/abs/https://academic.oup.com/mnras/article-pdf/207/1/23P/18223810/mnras207-023P.pdf}{{\ttfamily https://academic.oup.com/mnras/article-pdf/207/1/23P/18223810/mnras207-023P.pdf}}].

\bibitem{10.1143/PTP.72.1137}
Y.~Suto, K.~Sato and H.~Sato, \emph{Nonlinear evolution of negative density perturbations in a radiation-dominated universe}, \href{https://doi.org/10.1143/PTP.72.1137}{\emph{Progress of Theoretical Physics} {\bfseries 72} (1984) 1137} [\href{https://arxiv.org/abs/https://academic.oup.com/ptp/article-pdf/72/6/1137/5445109/72-6-1137.pdf}{{\ttfamily https://academic.oup.com/ptp/article-pdf/72/6/1137/5445109/72-6-1137.pdf}}].

\bibitem{1985GReGr..17..251S}
R.A.~{Sussman}, \emph{{Conformal structure of a Schwarzschild black hole immersed in a Friedmann universe.}}, \href{https://doi.org/10.1007/BF00760247}{\emph{General Relativity and Gravitation} {\bfseries 17} (1985) 251}.

\bibitem{Hellaby:1985zz}
C.~Hellaby and K.~Lake, \emph{{Shell crossings and the Tolman model}}, \href{https://doi.org/10.1086/162995}{\emph{Astrophys. J.} {\bfseries 290} (1985) 381}.

\bibitem{Hellaby1987}
C.~Hellaby, \emph{A kruskal-like model with finite density}, \href{https://doi.org/10.1088/0264-9381/4/3/021}{\emph{Classical and Quantum Gravity} {\bfseries 4} (1987) 635}.

\bibitem{PhysRevD.69.043502}
A.~Krasi\ifmmode~\acute{n}\else \'{n}\fi{}ski and C.~Hellaby, \emph{Formation of a galaxy with a central black hole in the lema\^{\i}tre-tolman model}, \href{https://doi.org/10.1103/PhysRevD.69.043502}{\emph{Phys. Rev. D} {\bfseries 69} (2004) 043502}.

\bibitem{Hellaby:2002nx}
C.~Hellaby and A.~Krasinski, \emph{{You can't get through Szekeres wormholes: Or, regularity, topology and causality in quasispherical Szekeres models}}, \href{https://doi.org/10.1103/PhysRevD.66.084011}{\emph{Phys. Rev. D} {\bfseries 66} (2002) 084011} [\href{https://arxiv.org/abs/gr-qc/0206052}{{\ttfamily gr-qc/0206052}}].

\bibitem{plebanski2006introduction}
J.~Plebanski and A.~Krasinski, \emph{An introduction to general relativity and cosmology}, Cambridge University Press (2006).

\bibitem{Gow_2023}
A.D.~Gow, H.~Assadullahi, J.H.P.~Jackson, K.~Koyama, V.~Vennin and D.~Wands, \emph{Non-perturbative non-gaussianity and primordial black holes}, \href{https://doi.org/10.1209/0295-5075/acd417}{\emph{Europhysics Letters} {\bfseries 142} (2023) 49001}.

\bibitem{Escriva:2023uko}
A.~Escriv\`a, V.~Atal and J.~Garriga, \emph{{Formation of trapped vacuum bubbles during inflation, and consequences for PBH scenarios}}, \href{https://doi.org/10.1088/1475-7516/2023/10/035}{\emph{JCAP} {\bfseries 10} (2023) 035} [\href{https://arxiv.org/abs/2306.09990}{{\ttfamily 2306.09990}}].

\bibitem{Shimada:2024eec}
M.~Shimada, A.~Escriv\'a, D.~Saito, K.~Uehara and C.-M.~Yoo, \emph{{Primordial Black Hole Formation from Type II Fluctuations with Primordial Non-Gaussianity}},  \href{https://arxiv.org/abs/2411.07648}{{\ttfamily 2411.07648}}.

\bibitem{Inui:2024fgk}
R.~Inui, C.~Joana, H.~Motohashi, S.~Pi, Y.~Tada and S.~Yokoyama, \emph{{Primordial black holes and induced gravitational waves from logarithmic non-Gaussianity}},  \href{https://arxiv.org/abs/2411.07647}{{\ttfamily 2411.07647}}.

\bibitem{Escriva:2025rja}
A.~Escriv\`a, \emph{{The threshold for PBH formation in the type-II region and its analytical estimation}},  \href{https://arxiv.org/abs/2504.05814}{{\ttfamily 2504.05814}}.

\bibitem{Deng:2016vzb}
H.~Deng, J.~Garriga and A.~Vilenkin, \emph{{Primordial black hole and wormhole formation by domain walls}}, \href{https://doi.org/10.1088/1475-7516/2017/04/050}{\emph{JCAP} {\bfseries 04} (2017) 050} [\href{https://arxiv.org/abs/1612.03753}{{\ttfamily 1612.03753}}].

\bibitem{Deng:2017uwc}
H.~Deng and A.~Vilenkin, \emph{{Primordial black hole formation by vacuum bubbles}}, \href{https://doi.org/10.1088/1475-7516/2017/12/044}{\emph{JCAP} {\bfseries 12} (2017) 044} [\href{https://arxiv.org/abs/1710.02865}{{\ttfamily 1710.02865}}].

\bibitem{Uehara:2024yyp}
K.~Uehara, A.~Escriv\`a, T.~Harada, D.~Saito and C.-M.~Yoo, \emph{{Numerical simulation of type II primordial black hole formation}}, \href{https://doi.org/10.1088/1475-7516/2025/01/003}{\emph{JCAP} {\bfseries 01} (2025) 003} [\href{https://arxiv.org/abs/2401.06329}{{\ttfamily 2401.06329}}].

\bibitem{Hayward:1993wb}
S.A.~Hayward, \emph{{General laws of black hole dynamics}}, \href{https://doi.org/10.1103/PhysRevD.49.6467}{\emph{Phys. Rev. D} {\bfseries 49} (1994) 6467}.

\bibitem{Hayward:1994bu}
S.A.~Hayward, \emph{{Gravitational energy in spherical symmetry}}, \href{https://doi.org/10.1103/PhysRevD.53.1938}{\emph{Phys. Rev. D} {\bfseries 53} (1996) 1938} [\href{https://arxiv.org/abs/gr-qc/9408002}{{\ttfamily gr-qc/9408002}}].

\bibitem{Maeda:2009tk}
H.~Maeda, T.~Harada and B.J.~Carr, \emph{{Cosmological wormholes}}, \href{https://doi.org/10.1103/PhysRevD.79.044034}{\emph{Phys. Rev. D} {\bfseries 79} (2009) 044034} [\href{https://arxiv.org/abs/0901.1153}{{\ttfamily 0901.1153}}].

\bibitem{Escriva:2025eqc}
A.~Escriv\`a, \emph{{A new approach for simulating PBH formation from generic curvature fluctuations with the Misner-Sharp formalism}},  \href{https://arxiv.org/abs/2504.05813}{{\ttfamily 2504.05813}}.

\bibitem{PhysRevD.60.084002}
M.~Shibata and M.~Sasaki, \emph{Black hole formation in the friedmann universe: Formulation and computation in numerical relativity}, \href{https://doi.org/10.1103/PhysRevD.60.084002}{\emph{Phys. Rev. D} {\bfseries 60} (1999) 084002}.

\bibitem{Lyth_2005}
D.H.~Lyth, K.A.~Malik and M.~Sasaki, \emph{A general proof of the conservation of the curvature perturbation}, \href{https://doi.org/10.1088/1475-7516/2005/05/004}{\emph{Journal of Cosmology and Astroparticle Physics} {\bfseries 2005} (2005) 004}.

\bibitem{Harada:2015ewt}
T.~Harada and S.~Jhingan, \emph{{Spherical and nonspherical models of primordial black hole formation: exact solutions}}, \href{https://doi.org/10.1093/ptep/ptw123}{\emph{PTEP} {\bfseries 2016} (2016) 093E04} [\href{https://arxiv.org/abs/1512.08639}{{\ttfamily 1512.08639}}].

\bibitem{Tolman:1934za}
R.C.~Tolman, \emph{{Effect of imhomogeneity on cosmological models}}, \href{https://doi.org/10.1073/pnas.20.3.169}{\emph{Proc. Nat. Acad. Sci.} {\bfseries 20} (1934) 169}.

\bibitem{Polnarev:2006aa}
A.G.~Polnarev and I.~Musco, \emph{{Curvature profiles as initial conditions for primordial black hole formation}}, \href{https://doi.org/10.1088/0264-9381/24/6/003}{\emph{Class. Quant. Grav.} {\bfseries 24} (2007) 1405} [\href{https://arxiv.org/abs/gr-qc/0605122}{{\ttfamily gr-qc/0605122}}].

\bibitem{PhysRevD.91.084057}
T.~Harada, C.-M.~Yoo, T.~Nakama and Y.~Koga, \emph{Cosmological long-wavelength solutions and primordial black hole formation}, \href{https://doi.org/10.1103/PhysRevD.91.084057}{\emph{Phys. Rev. D} {\bfseries 91} (2015) 084057}.

\bibitem{Lemaitre:1933gd}
G.~Lemaitre, \emph{{The expanding universe}}, \href{https://doi.org/10.1023/A:1018855621348}{\emph{Annales Soc. Sci. Bruxelles A} {\bfseries 53} (1933) 51}.

\bibitem{Misner:1973prb}
C.W.~Misner, K.S.~Thorne and J.A.~Wheeler, \emph{{Gravitation}}, W. H. Freeman, San Francisco (1973).

\bibitem{galaxies10060112}
C.-M.~Yoo, \emph{The basics of primordial black hole formation and abundance estimation}, \href{https://doi.org/10.3390/galaxies10060112}{\emph{Galaxies} {\bfseries 10} (2022) }.

\bibitem{1977A&A....59...53S}
J.~{Silk}, \emph{{Large-scale inhomogeneity of the universe: spherically symmetric models.}}, {\emph{\aap} {\bfseries 59} (1977) 53}.

\bibitem{PhysRevD.49.6467}
S.A.~Hayward, \emph{General laws of black-hole dynamics}, \href{https://doi.org/10.1103/PhysRevD.49.6467}{\emph{Phys. Rev. D} {\bfseries 49} (1994) 6467}.

\bibitem{BA67359925}
E.~Poisson, \emph{A relativist's toolkit : the mathematics of black-hole mechanics}, Cambridge University Press (2004).

\bibitem{Zeldovich:1983cr}
Y.B.~Zeldovich and I.D.~Novikov, \emph{{RELATIVISTIC ASTROPHYSICS. VOL. 2. THE STRUCTURE AND EVOLUTION OF THE UNIVERSE}} (1983).

\bibitem{Novikov:1989sz}
I.D.~Novikov and V.P.~Frolov, \emph{{PHYSICS OF BLACK HOLES}}, Kluwer Academic, Dordrecht, Netherlands (1989), \href{https://doi.org/10.1007/978-94-017-2651-1}{10.1007/978-94-017-2651-1}.

\bibitem{Frolov:1998wf}
V.P.~Frolov and I.D.~Novikov, eds., \emph{{Black hole physics: Basic concepts and new developments}} (1998), \href{https://doi.org/10.1007/978-94-011-5139-9}{10.1007/978-94-011-5139-9}.

\bibitem{1963JETP...16..732Z}
Y.B.~{Zel'Dovich}, \emph{{Semiclosed Worlds in the General Theory of Relativity}}, {\emph{Soviet Journal of Experimental and Theoretical Physics} {\bfseries 16} (1963) 732}.

\bibitem{1964SvA.....7..587N}
I.D.~{Novikov}, \emph{{On the Evolution of a Semiclosed World}}, {\emph{\sovast} {\bfseries 7} (1964) 587}.

\bibitem{gourgoulhonBH}
E.~Gourgoulhon, \emph{Geometry and physics of black holes},  2022.
\newblock \url{https://luth.obspm.fr/~luthier/gourgoulhon/bh16/}.

\bibitem{Yoo:2024lhp}
C.-M.~Yoo, \emph{{Primordial black hole formation from a nonspherical density profile with a misaligned deformation tensor}}, \href{https://doi.org/10.1103/PhysRevD.110.043526}{\emph{Phys. Rev. D} {\bfseries 110} (2024) 043526} [\href{https://arxiv.org/abs/2403.11147}{{\ttfamily 2403.11147}}].

\bibitem{PhysRevD.105.103538}
C.-M.~Yoo, T.~Harada, S.~Hirano, H.~Okawa and M.~Sasaki, \emph{Primordial black hole formation from massless scalar isocurvature}, \href{https://doi.org/10.1103/PhysRevD.105.103538}{\emph{Phys. Rev. D} {\bfseries 105} (2022) 103538}.

\bibitem{PhysRevLett.111.161102}
C.-M.~Yoo, H.~Okawa and K.-i.~Nakao, \emph{Black-hole universe: Time evolution}, \href{https://doi.org/10.1103/PhysRevLett.111.161102}{\emph{Phys. Rev. Lett.} {\bfseries 111} (2013) 161102}.

\bibitem{PhysRevD.89.104032}
H.~Okawa, H.~Witek and V.~Cardoso, \emph{Black holes and fundamental fields in numerical relativity: Initial data construction and evolution of bound states}, \href{https://doi.org/10.1103/PhysRevD.89.104032}{\emph{Phys. Rev. D} {\bfseries 89} (2014) 104032}.

\bibitem{Musco:2012au}
I.~Musco and J.C.~Miller, \emph{{Primordial black hole formation in the early universe: critical behaviour and self-similarity}}, \href{https://doi.org/10.1088/0264-9381/30/14/145009}{\emph{Class. Quant. Grav.} {\bfseries 30} (2013) 145009} [\href{https://arxiv.org/abs/1201.2379}{{\ttfamily 1201.2379}}].

\bibitem{Escriva:2020tak}
A.~Escriv\`a, C.~Germani and R.K.~Sheth, \emph{{Analytical thresholds for black hole formation in general cosmological backgrounds}}, \href{https://doi.org/10.1088/1475-7516/2021/01/030}{\emph{JCAP} {\bfseries 01} (2021) 030} [\href{https://arxiv.org/abs/2007.05564}{{\ttfamily 2007.05564}}].

\end{thebibliography}\endgroup
\end{document}